\documentclass[12pt]{article}%
\usepackage{amssymb}
\usepackage{amsmath}
\usepackage{amsfonts}
\usepackage{graphicx}%
\newenvironment{proof}[1][Proof]{\noindent\textbf{#1.} }{\ \rule{0.5em}{0.5em}}
\begin{document}

\author{{\small V. V. Fern\'{a}ndez}$^{{\footnotesize 1}}${\small , A. M.
Moya}$^{{\footnotesize 1}}${\small , E. Notte-Cuello}$^{{\footnotesize 2}}%
${\small \ and W. A. Rodrigues Jr.}$^{{\footnotesize 1}}${\small . }\\$^{{\footnotesize 1}}\hspace{-0.1cm}${\footnotesize Institute of Mathematics,
Statistics and Scientific Computation}\\{\footnotesize IMECC-UNICAMP CP 6065}\\{\footnotesize 13083-859 Campinas, SP, Brazil}\\$^{{\footnotesize 2}}${\small Departamento de} {\small Matem\'{a}ticas,}\\{\small Universidad de La Serena}\\{\small Av. Cisternas 1200, La Serena-Chile}\\{\small e-mail:} {\small walrod@ime.unicamp.br and enotte@userena.cl }}
\title{Covariant Derivatives of Extensor Fields}
\maketitle

\begin{abstract}
A simple theory of the covariant derivatives, deformed derivatives and
relative covariant derivatives of \textit{extensor} fields is present using
algebraic and analytical tools developed in previous papers. Several important
formulas are derived.

\end{abstract}

\newpage

\section{Introduction}

A simple theory of the covariant derivatives, deformed derivatives and
relative covariant derivatives for $%
%TCIMACRO{\dbigwedge }%
%BeginExpansion
{\displaystyle\bigwedge}
%EndExpansion
TM$ and $%
%TCIMACRO{\dbigwedge }%
%BeginExpansion
{\displaystyle\bigwedge}
%EndExpansion
T^{\ast}M $ valued \textit{extensor} fields on a arbitrary smooth manifold
$M$, which are important in, e.g., intrinsic formulations of geometrical
theories of the gravitational field and in the Lagrangian formalism of field
theories \cite{rodoliv2006} is presented using algebraic and analytical tools
developed in previous papers \cite{fmcr1,fmcr3,fmcr4}. Useful properties of
these concepts are present with details.

\section{Extensor Fields}

Let $U$ be an open set on the finite dimensional smooth manifold $M$ (i.e.,
$\dim M=n$ with $n\in\mathbb{N}$). As usual, the ring (with identity) of the
smooth scalar fields on $U$ will denoted by $\mathcal{S}(U)$. The module over
$\mathcal{S}(U)$ of the smooth vector fields on $U$ will be symbolized as
$\mathcal{V}(U).$

The modules over $\mathcal{S}(U)$ of the smooth\footnote{In this paper, smooth
menas $\mathcal{C}^{\infty}$-differentiable, or at least enough differentiable
for our statements to hold.} multivector fields on $U$ and the smooth
multiform fields on $U$ will be respectively denoted by $%
%TCIMACRO{\dbigwedge }%
%BeginExpansion
{\displaystyle\bigwedge}
%EndExpansion
\mathcal{V}(U)$ and $%
%TCIMACRO{\dbigwedge }%
%BeginExpansion
{\displaystyle\bigwedge}
%EndExpansion
\mathcal{V}^{\ast}(U).$

A multivector extensor mapping
\begin{equation}
\tau:U\longrightarrow\underset{p\in U}{%
%TCIMACRO{\tbigcup }%
%BeginExpansion
{\textstyle\bigcup}
%EndExpansion
}\left.  \overset{\left.  {}\right.  }{ext}\right.  _{k}^{l}(T_{p}M)
\end{equation}
such that for each $p\in U$, $\tau_{(p)}\in\left.  \overset{\left.  {}\right.
}{ext}\right.  _{k}^{l}(T_{p}M)$ is called a\emph{\ multivector extensor field
of }$k$\emph{\ multivector and }$l$\emph{\ multiform variables on }$U.$

A multiform extensor mapping
\begin{equation}
\upsilon:U\longrightarrow\underset{p\in U}{%
%TCIMACRO{\tbigcup }%
%BeginExpansion
{\textstyle\bigcup}
%EndExpansion
}\left.  \overset{\ast}{ext}\right.  _{k}^{l}(T_{p}M)
\end{equation}
such that for each $p\in U$, $\upsilon_{(p)}\in\left.  \overset{\ast}%
{ext}\right.  _{k}^{l}(T_{p}M)$ is called a\emph{\ multiform }%
\textit{extensor}\emph{\textbf{\ }field of }$k$\emph{\ multivector and }%
$l$\emph{\ multiform variables on }$U.$

In the above formulas, $\left.  \overset{\left.  {}\right.  }{ext}\right.
_{k}^{l}(T_{p}M)$ is a short notation for the space of multivector extensors
of $k$ multivector and $l$ multiform variables over $T_{p}M,$ i.e., for each
$p\in U:$%
\[
\left.  \overset{\left.  {}\right.  }{ext}\right.  _{k}^{l}(T_{p}M):=ext(%
%TCIMACRO{\dbigwedge \nolimits_{1}^{\Diamond}}%
%BeginExpansion
{\displaystyle\bigwedge\nolimits_{1}^{\Diamond}}
%EndExpansion
T_{p}M,\ldots,%
%TCIMACRO{\dbigwedge \nolimits_{k}^{\Diamond}}%
%BeginExpansion
{\displaystyle\bigwedge\nolimits_{k}^{\Diamond}}
%EndExpansion
T_{p}M,%
%TCIMACRO{\dbigwedge \nolimits_{1}^{\Diamond}}%
%BeginExpansion
{\displaystyle\bigwedge\nolimits_{1}^{\Diamond}}
%EndExpansion
T_{p}^{\star}M,\ldots,%
%TCIMACRO{\dbigwedge \nolimits_{l}^{\Diamond}}%
%BeginExpansion
{\displaystyle\bigwedge\nolimits_{l}^{\Diamond}}
%EndExpansion
T_{p}^{\star}M;%
%TCIMACRO{\dbigwedge \nolimits^{\Diamond}}%
%BeginExpansion
{\displaystyle\bigwedge\nolimits^{\Diamond}}
%EndExpansion
T_{p}M),
\]
and $\left.  \overset{\ast}{ext}\right.  _{k}^{l}(T_{p}M)$ is a shorth
notation for the space of multiform \textit{extensors} of $k$ multivector and
$l$ multiform variables over $T_{p}M,$ i.e., for each $p\in U:$%
\[
\left.  \overset{\ast}{ext}\right.  _{k}^{l}(T_{p}M):=ext(%
%TCIMACRO{\dbigwedge \nolimits_{1}^{\Diamond}}%
%BeginExpansion
{\displaystyle\bigwedge\nolimits_{1}^{\Diamond}}
%EndExpansion
T_{p}M,\ldots,%
%TCIMACRO{\dbigwedge \nolimits_{k}^{\Diamond}}%
%BeginExpansion
{\displaystyle\bigwedge\nolimits_{k}^{\Diamond}}
%EndExpansion
T_{p}M,%
%TCIMACRO{\dbigwedge \nolimits_{1}^{\Diamond}}%
%BeginExpansion
{\displaystyle\bigwedge\nolimits_{1}^{\Diamond}}
%EndExpansion
T_{p}^{\star}M,\ldots,%
%TCIMACRO{\dbigwedge \nolimits_{l}^{\Diamond}}%
%BeginExpansion
{\displaystyle\bigwedge\nolimits_{l}^{\Diamond}}
%EndExpansion
T_{p}^{\star}M;%
%TCIMACRO{\dbigwedge \nolimits^{\Diamond}}%
%BeginExpansion
{\displaystyle\bigwedge\nolimits^{\Diamond}}
%EndExpansion
T_{p}^{\star}M).
\]

Let us denote the smooth multivector fields: $U\ni p\longmapsto X_{1(p)}\in%
%TCIMACRO{\dbigwedge \nolimits_{1}^{\Diamond}}%
%BeginExpansion
{\displaystyle\bigwedge\nolimits_{1}^{\Diamond}}
%EndExpansion
T_{p}M,\ldots$, $U\ni p\longmapsto X_{k(p)}\in%
%TCIMACRO{\dbigwedge \nolimits_{k}^{\Diamond}}%
%BeginExpansion
{\displaystyle\bigwedge\nolimits_{k}^{\Diamond}}
%EndExpansion
T_{p}M$, and $U\ni p\longmapsto X_{(p)}\in%
%TCIMACRO{\dbigwedge \nolimits^{\Diamond}}%
%BeginExpansion
{\displaystyle\bigwedge\nolimits^{\Diamond}}
%EndExpansion
T_{p}M$, respectively by $%
%TCIMACRO{\dbigwedge \nolimits_{1}^{\Diamond}}%
%BeginExpansion
{\displaystyle\bigwedge\nolimits_{1}^{\Diamond}}
%EndExpansion
\mathcal{V}(U),\ldots$, $%
%TCIMACRO{\dbigwedge \nolimits_{k}^{\Diamond}}%
%BeginExpansion
{\displaystyle\bigwedge\nolimits_{k}^{\Diamond}}
%EndExpansion
\mathcal{V}(U)$, and $%
%TCIMACRO{\dbigwedge \nolimits^{\Diamond}}%
%BeginExpansion
{\displaystyle\bigwedge\nolimits^{\Diamond}}
%EndExpansion
\mathcal{V}(U)$. Let us denote the smooth multiform fields: $U\ni
p\longmapsto\Phi_{(p)}^{1}\in%
%TCIMACRO{\dbigwedge \nolimits_{1}^{\Diamond}}%
%BeginExpansion
{\displaystyle\bigwedge\nolimits_{1}^{\Diamond}}
%EndExpansion
T_{p}^{\ast}M,\ldots$, $U\ni p\longmapsto\Phi_{(p)}^{l}\in%
%TCIMACRO{\dbigwedge \nolimits_{l}^{\Diamond}}%
%BeginExpansion
{\displaystyle\bigwedge\nolimits_{l}^{\Diamond}}
%EndExpansion
T_{p}^{\ast}M$, and $U\ni p\longmapsto\Phi_{(p)}^{l}\in%
%TCIMACRO{\dbigwedge \nolimits^{\Diamond}}%
%BeginExpansion
{\displaystyle\bigwedge\nolimits^{\Diamond}}
%EndExpansion
T_{p}^{\ast}M$, by $%
%TCIMACRO{\dbigwedge \nolimits_{1}^{\Diamond}}%
%BeginExpansion
{\displaystyle\bigwedge\nolimits_{1}^{\Diamond}}
%EndExpansion
\mathcal{V}^{\ast}(U),$ $\ldots,$ $%
%TCIMACRO{\dbigwedge \nolimits_{l}^{\Diamond}}%
%BeginExpansion
{\displaystyle\bigwedge\nolimits_{l}^{\Diamond}}
%EndExpansion
\mathcal{V}^{\ast}(U)$ and $%
%TCIMACRO{\dbigwedge \nolimits^{\Diamond}}%
%BeginExpansion
{\displaystyle\bigwedge\nolimits^{\Diamond}}
%EndExpansion
\mathcal{V}^{\ast}(U)$.

Such a multivector extensor field $\tau$ will be said to be smooth, if and
only if, for all $X_{1}\in%
%TCIMACRO{\dbigwedge \nolimits_{1}^{\Diamond}}%
%BeginExpansion
{\displaystyle\bigwedge\nolimits_{1}^{\Diamond}}
%EndExpansion
\mathcal{V}(U),\ldots,X_{k}\in%
%TCIMACRO{\dbigwedge \nolimits_{k}^{\Diamond}}%
%BeginExpansion
{\displaystyle\bigwedge\nolimits_{k}^{\Diamond}}
%EndExpansion
\mathcal{V}(U),$ and for all $\Phi^{1}\in%
%TCIMACRO{\dbigwedge \nolimits_{1}^{\Diamond}}%
%BeginExpansion
{\displaystyle\bigwedge\nolimits_{1}^{\Diamond}}
%EndExpansion
\mathcal{V}^{\ast}(U),\ldots,\Phi^{l}\in%
%TCIMACRO{\dbigwedge \nolimits_{l}^{\Diamond}}%
%BeginExpansion
{\displaystyle\bigwedge\nolimits_{l}^{\Diamond}}
%EndExpansion
\mathcal{V}^{\star}(U)$, the multivector mapping defined by
\begin{equation}
U\ni p\longmapsto\tau_{(p)}(X_{1(p)},\ldots,X_{k(p)},\Phi_{(p)}^{1}%
,\ldots,\Phi_{(p)}^{l})\in%
%TCIMACRO{\dbigwedge \nolimits^{\Diamond}}%
%BeginExpansion
{\displaystyle\bigwedge\nolimits^{\Diamond}}
%EndExpansion
T_{p}M \label{EF3}%
\end{equation}
is a smooth multivector field on $U,$ (i.e., an object living on $%
%TCIMACRO{\dbigwedge \nolimits^{\Diamond}}%
%BeginExpansion
{\displaystyle\bigwedge\nolimits^{\Diamond}}
%EndExpansion
\mathcal{V}(U)$).

Such a multiform extensor field $\upsilon$ will be said to be smooth , if and
only if, for all $X_{1}\in%
%TCIMACRO{\dbigwedge \nolimits_{1}^{\Diamond}}%
%BeginExpansion
{\displaystyle\bigwedge\nolimits_{1}^{\Diamond}}
%EndExpansion
\mathcal{V}(U),\ldots,X_{k}\in%
%TCIMACRO{\dbigwedge \nolimits_{k}^{\Diamond}}%
%BeginExpansion
{\displaystyle\bigwedge\nolimits_{k}^{\Diamond}}
%EndExpansion
\mathcal{V}(U),$ and for all $\Phi^{1}\in%
%TCIMACRO{\dbigwedge \nolimits_{1}^{\Diamond}}%
%BeginExpansion
{\displaystyle\bigwedge\nolimits_{1}^{\Diamond}}
%EndExpansion
\mathcal{V}^{\ast}(U),\ldots,\Phi^{l}\in%
%TCIMACRO{\dbigwedge \nolimits_{l}^{\Diamond}}%
%BeginExpansion
{\displaystyle\bigwedge\nolimits_{l}^{\Diamond}}
%EndExpansion
\mathcal{V}^{\ast}(U)$, the multiform mapping defined by
\begin{equation}
U\ni p\longmapsto\upsilon_{(p)}(X_{1(p)},\ldots,X_{k(p)},\Phi_{(p)}^{1}%
,\ldots,\Phi_{(p)}^{l})\in%
%TCIMACRO{\dbigwedge \nolimits^{\Diamond}}%
%BeginExpansion
{\displaystyle\bigwedge\nolimits^{\Diamond}}
%EndExpansion
T_{p}^{\star}M \label{EF4}%
\end{equation}
is a smooth multiform field on $U,$ (i.e., an object living on $%
%TCIMACRO{\dbigwedge ^{\Diamond}}%
%BeginExpansion
{\displaystyle\bigwedge^{\Diamond}}
%EndExpansion
\mathcal{V}^{\star}(U)$).

We emphasize\footnote{A \emph{short name} for a multivector (or, multiform)
extensor of $k$ multivector and $l$ multiform variables could be: a $\binom
{l}{k}$ multivector (respectively, multiform) extensor.} that according with
the definitions of smoothness as given above, a smooth\emph{\ }$\binom{l}{k}$
multivector extensor field on $U$ \emph{can be identified} to a $\binom{l}{k}$
multivector extensor over $\mathcal{V}(U).$ It is also true that a
smooth\emph{\ }$\binom{l}{k}$ multiform extensor field on $U$ \emph{can be
properly seen }as a $\binom{l}{k}$ multiform extensor over $\mathcal{V}(U).$

Thus, the set of smooth\emph{\ }$\binom{l}{k}$ multivector extensor fields on
$U$ is just a module over $\mathcal{S}(U)$ which could be denoted by $\left.
\overset{\left.  {}\right.  }{ext}\right.  _{k}^{l}\mathcal{V}(U).$ And, the
set of smooth\emph{\ }$\binom{l}{k}$ multiform extensor fields on $U$ is also
a module over $\mathcal{S}(U)$ which can be symbolized as $\left.
\overset{\ast}{ext}\right.  _{k}^{l}\mathcal{V}(U).$

\subsection{Algebras of Extensor Fields}

We define now the exterior products of smooth multivector extensor fields on
$U$ and smooth multiform extensor\textbf{\ }fields on $U$. We also present the
definitions of smooth multivector extensor\textbf{\ }fields on $U$ with smooth
multiform extensor\textbf{\ }fields on $U.$

The exterior product of either multivector extensor fields or multiform
extensor fields $\tau$ and $\sigma$ is defined as%
\begin{equation}
\left(  \tau\wedge\sigma\right)  _{(p)}=\tau_{(p)}\wedge\sigma_{(p)}
\label{EF5}%
\end{equation}
for every $p\in U$.

Each module over $\mathcal{S}(U)$ of either the smooth multivector extensor
fields on $U$ or the smooth multiform extensor fields on $U$ endowed with the
respective exterior product is an associative algebra.

The duality scalar product of a multiform extensor field $\tau$ with a
multivector extensor field $\sigma$ is a scalar extensor field $\left\langle
\tau,\sigma\right\rangle $ defined by%
\begin{equation}
\left\langle \tau,\sigma\right\rangle _{(p)}=\left\langle \tau_{(p)}%
,\sigma_{(p)}\right\rangle , \label{EF6}%
\end{equation}
for every $p\in U.$

The duality left contracted product of a multiform extensor field $\tau$ with
a multivector extensor field $\sigma$ (or, a multivector extensor field $\tau$
with a multiform extensor field $\sigma$) is the multivector extensor field
(respectively, the multiform extensor field) denoted by $\left\langle
\tau,\sigma\right\vert $ and defined by%
\begin{equation}
\left\langle \tau,\sigma\right\vert _{(p)}=\left\langle \tau_{(p)}%
,\sigma_{(p)}\right\vert , \label{EF7}%
\end{equation}
for every $p\in U.$

The duality right contracted product of a multiform extensor field $\tau$ with
a multivector extensor field $\sigma$ (or, a multivector extensor field $\tau$
with a multiform extensor field $\sigma$) is the multiform extensor field
(respectively, the multivector extensor field) named as $\left\vert
\tau,\sigma\right\rangle $, and defined by%
\begin{equation}
\left\vert \tau,\sigma\right\rangle _{(p)}=\left\vert \tau_{(p)},\sigma
_{(p)}\right\rangle \label{EF8}%
\end{equation}
for every $p\in U.$

Each duality contracted product of smooth multivector extensor fields on $U$
with smooth multiform fields on $U$ yields a non-associative algebra.

\section{Covariant Derivative of Extensor Fields}

Let $\left\langle U,\Gamma\right\rangle $ be a parallelism structure
\cite{fmcr3} on $U,$ and let us take $a\in\mathcal{V}(U)$. The $a$%
\emph{-Directional Covariant Derivatives }($a$\emph{-DCD})\emph{,} associated
with $\left\langle U,\Gamma\right\rangle $,\emph{\ }of a smooth
\textit{multivector} extensor field on $U$ or a smooth \textit{multiform}
extensor field on $U$ are the mappings
\[
\left.  \overset{\left.  {}\right.  }{ext}\right.  _{k}^{l}\mathcal{V}%
(U)\ni\tau\longmapsto\nabla_{a}\tau\in\left.  \overset{\left.  {}\right.
}{ext}\right.  _{k}^{l}\mathcal{V}(U),
\]
and
\[
\left.  \overset{\ast}{ext}\right.  _{k}^{l}\mathcal{V}(U)\ni\tau
\longmapsto\nabla_{a}\tau\in\left.  \overset{\ast}{ext}\right.  _{k}%
^{l}\mathcal{V}(U),
\]
such that for all $X_{1}\in%
%TCIMACRO{\dbigwedge \nolimits_{1}^{\Diamond}}%
%BeginExpansion
{\displaystyle\bigwedge\nolimits_{1}^{\Diamond}}
%EndExpansion
\mathcal{V}(U),\ldots,X_{k}\in%
%TCIMACRO{\dbigwedge \nolimits_{k}^{\Diamond}}%
%BeginExpansion
{\displaystyle\bigwedge\nolimits_{k}^{\Diamond}}
%EndExpansion
\mathcal{V}(U),$ and for all $\Phi^{1}\in%
%TCIMACRO{\dbigwedge \nolimits_{1}^{\Diamond}}%
%BeginExpansion
{\displaystyle\bigwedge\nolimits_{1}^{\Diamond}}
%EndExpansion
\mathcal{V}^{\ast}(U),\ldots,$ $\Phi^{l}\in%
%TCIMACRO{\dbigwedge \nolimits_{l}^{\Diamond}}%
%BeginExpansion
{\displaystyle\bigwedge\nolimits_{l}^{\Diamond}}
%EndExpansion
\mathcal{V}^{\ast}(U)$ we have%
\begin{align}
\nabla_{a}\tau(X_{1},...,X_{k},\Phi^{1},...,\Phi^{l})  &  =\nabla_{a}%
(\tau(X_{1},...,X_{k},\Phi^{1},...,\Phi^{l}))\nonumber\\
&  -\tau(\nabla_{a}X_{1},...,X_{k},\Phi^{1},...,\Phi^{l})-\cdots\nonumber\\
&  -\tau(X_{1},...,\nabla_{a}X_{k},\Phi^{1},...,\Phi^{l})\nonumber\\
&  -\tau(X_{1},...,X_{k},\nabla_{a}\Phi^{1},...,\Phi^{l})-\cdots\nonumber\\
&  -\tau(X_{1},...,X_{k},\Phi^{1},...,\nabla_{a}\Phi^{l}). \label{CDEF1}%
\end{align}

The covariant derivative of smooth multivector (or multiform) extensor fields
has two basic properties.

\begin{itemize}
\item For $f\in\mathcal{S}(U),$ and $a,b\in\mathcal{V}(U),\mathcal{\ }$and
$\tau\in\left.  \overset{\left.  {}\right.  }{ext}\right.  _{k}^{l}%
\mathcal{V}(U)$ (or $\tau\in\left.  \overset{\ast}{ext}\right.  _{k}%
^{l}\mathcal{V}(U)$)%
\begin{align}
\nabla_{a+b}\tau &  =\nabla_{a}\tau+\nabla_{b}\tau\label{CDEF2a}\\
\nabla_{fa}\tau &  =f\nabla_{a}\tau. \label{CDEF2b}%
\end{align}

\item For $f\in\mathcal{S}(U),$ and $a\in\mathcal{V}(U),\mathcal{\ }$and
$\tau,\sigma\in\left.  \overset{\left.  {}\right.  }{ext}\right.  _{k}%
^{l}\mathcal{V}(U)$ (or $\tau,\sigma\in\left.  \overset{\ast}{ext}\right.
_{k}^{l}\mathcal{V}(U)$)%
\begin{align}
\nabla_{a}(\tau+\sigma)  &  =\nabla_{a}\tau+\nabla_{a}\sigma,\label{CDEF3a}\\
\nabla_{a}(f\tau)  &  =(af)\tau+f\nabla_{a}\tau. \label{CDEF3b}%
\end{align}

\end{itemize}

The covariant differentiation of the exterior product of smooth multivector
(or multiform) extensor fields satisfies the Leibniz's rule.

\begin{itemize}
\item For all $\tau\in\left.  \overset{\left.  {}\right.  }{ext}\right.
_{k}^{l}\mathcal{V}(U)$ and $\sigma\in\left.  \overset{\left.  {}\right.
}{ext}\right.  _{r}^{s}\mathcal{V}(U)$ (or, $\tau\in\left.  \overset{\ast
}{ext}\right.  _{k}^{l}\mathcal{V}(U)$ and $\sigma\in\left.  \overset{\ast
}{ext}\right.  _{r}^{s}\mathcal{V}(U)$), it holds
\begin{equation}
\nabla_{a}(\tau\wedge\sigma)=(\nabla_{a}\tau)\wedge\sigma+\tau\wedge\left(
\nabla_{a}\sigma\right)  . \label{CDEF4}%
\end{equation}

\end{itemize}

\begin{proof}
Without loss of generality, we prove this statement only for multivector
extensor fields $(X,\Phi)\longmapsto\tau(X,\Phi)$ and $(Y,\Psi)\longmapsto
\sigma(Y,\Psi).$ Using Eq.(\ref{CDEF1}), we can write
\begin{align*}
&  \nabla_{a}(\tau\wedge\sigma)(X,Y,\Phi,\Psi)\\
&  =\nabla_{a}((\tau\wedge\sigma)(X,Y,\Phi,\Psi))\\
&  -(\tau\wedge\sigma)(\nabla_{a}X,Y,\Phi,\Psi)-(\tau\wedge\sigma
)(X,\nabla_{a}Y,\Phi,\Psi)\\
&  -(\tau\wedge\sigma)(X,Y,\nabla_{a}\Phi,\Psi)-(\tau\wedge\sigma
)(X,Y,\Phi,\nabla_{a}\Psi).
\end{align*}

Using Eq.(\ref{EF5}) and recalling Leibniz's rule for the covariant
differentiation of the exterior product of multivector fields, we have
\begin{align*}
&  \nabla_{a}(\tau\wedge\sigma)(X,Y,\Phi,\Psi)\\
&  =\nabla_{a}(\tau(X,\Phi))\wedge\sigma(Y,\Psi)+\tau(X,\Phi)\wedge\nabla
_{a}(\sigma(Y,\Psi))\\
&  -\tau(\nabla_{a}X,\Phi)\wedge\sigma(Y,\Psi)-\tau(X,\Phi)\wedge\sigma
(\nabla_{a}Y,\Psi)\\
&  -\tau(X,\nabla_{a}\Phi)\wedge\sigma(Y,\Psi)-\tau(X,\Phi)\wedge
\sigma(Y,\nabla_{a}\Psi),
\end{align*}
i.e.,
\begin{align*}
&  \nabla_{a}(\tau\wedge\sigma)(X,Y,\Phi,\Psi)\\
&  =(\nabla_{a}(\tau(X,\Phi))-\tau(\nabla_{a}X,\Phi)-\tau(X,\nabla_{a}%
\Phi))\wedge\sigma(Y,\Psi)\\
&  +\tau(X,\Phi)\wedge(\nabla_{a}(\sigma(Y,\Psi))-\sigma(\nabla_{a}%
Y,\Psi)-\sigma(Y,\nabla_{a}\Psi)).
\end{align*}
Then, using once again Eq.(\ref{CDEF1}) and Eq.(\ref{EF5}), the expected
result follows.
\end{proof}

The covariant differentiation of the duality scalar product of smooth extensor
fields satisfies the Leibniz's rule.

\begin{itemize}
\item For all $\tau\in\left.  \overset{\ast}{ext}\right.  _{k}^{l}%
\mathcal{V}(U)$ and $\sigma\in\left.  \overset{\left.  {}\right.  }%
{ext}\right.  _{r}^{s}\mathcal{V}(U)$ (or, $\tau\in\left.  \overset{\left.
{}\right.  }{ext}\right.  _{k}^{l}\mathcal{V}(U)$ and $\tau\in\left.
\overset{\ast}{ext}\right.  _{r}^{s}\mathcal{V}(U)$), we have that
\begin{equation}
\nabla_{a}\left\langle \tau,\sigma\right\rangle =\left\langle \nabla_{a}%
\tau,\sigma\right\rangle +\left\langle \tau,\nabla_{a}\sigma\right\rangle .
\label{CDEF5}%
\end{equation}

\end{itemize}

\begin{proof}
Without loss of generality, we prove this statement only for a multiform
extensor field $\tau$ and a multivector extensor field $\sigma$ such that
$(X,\Phi)\longmapsto\tau(X,\Phi)$ and $(Y,\Psi)\longmapsto\sigma(Y,\Psi).$
Using Eq.(\textbf{\ref{CDEF1}}), we can write%
\begin{align*}
&  \nabla_{a}\left\langle \tau,\sigma\right\rangle (X,Y,\Phi,\Psi)\\
&  =\nabla_{a}(\left\langle \tau,\sigma\right\rangle (X,Y,\Phi,\Psi))\\
&  -\left\langle \tau,\sigma\right\rangle (\nabla_{a}X,Y,\Phi,\Psi
)-\left\langle \tau,\sigma\right\rangle (X,\nabla_{a}Y,\Phi,\Psi)\\
&  -\left\langle \tau,\sigma\right\rangle (X,Y,\nabla_{a}\Phi,\Psi
)-\left\langle \tau,\sigma\right\rangle (X,Y,\Phi,\nabla_{a}\Psi).
\end{align*}
Next, using Eq.(\textbf{\ref{EF6}}) and recalling Leibniz's rule for the
covariant differentiation of the duality scalar product of multiform fields
with multivector fields, we have
\begin{align*}
&  \nabla_{a}\left\langle \tau,\sigma\right\rangle (X,Y,\Phi,\Psi)\\
&  =\left\langle \nabla_{a}(\tau(X,\Phi)),\sigma(Y,\Psi)\right\rangle
+\left\langle \tau(X,\Phi),\nabla_{a}(\sigma(Y,\Psi))\right\rangle \\
&  -\left\langle \tau(\nabla_{a}X,\Phi),\sigma(Y,\Psi)\right\rangle
-\left\langle \tau(X,\Phi),\sigma(\nabla_{a}Y,\Psi)\right\rangle \\
&  -\left\langle \tau(X,\nabla_{a}\Phi),\sigma(Y,\Psi)\right\rangle
-\left\langle \tau(X,\Phi),\sigma(Y,\nabla_{a}\Psi)\right\rangle ,
\end{align*}
i.e.,
\begin{align*}
&  \nabla_{a}\left\langle \tau,\sigma\right\rangle (X,Y,\Phi,\Psi)\\
&  =\left\langle (\nabla_{a}(\tau(X,\Phi))-\tau(\nabla_{a}X,\Phi
)-\tau(X,\nabla_{a}\Phi)),\sigma(Y,\Psi)\right\rangle \\
&  +\left\langle \tau(X,\Phi),(\nabla_{a}(\sigma(Y,\Psi))-\sigma(\nabla
_{a}Y,\Psi)-\sigma(Y,\nabla_{a}\Psi))\right\rangle .
\end{align*}
Then, using once again Eq.(\ref{CDEF1}) and Eq.(\ref{EF6}), the expected
result follows.
\end{proof}

The covariant differentiation of each one of the duality contracted products
of smooth extensor fields satisfies the Leibniz's rule.

\begin{itemize}
\item For all $\tau\in\left.  \overset{\ast}{ext}\right.  _{k}^{l}%
\mathcal{V}(U)$ and $\sigma\in\left.  \overset{\left.  {}\right.  }%
{ext}\right.  _{r}^{s}\mathcal{V}(U)$ (or, $\tau\in\left.  \overset{\left.
{}\right.  }{ext}\right.  _{k}^{l}\mathcal{V}(U)$ and $\tau\in\left.
\overset{\ast}{ext}\right.  _{r}^{s}\mathcal{V}(U)$), it holds
\begin{align}
\nabla_{a}\left\langle \tau,\sigma\right\vert  &  =\left\langle \nabla_{a}%
\tau,\sigma\right\vert +\left\langle \tau,\nabla_{a}\sigma\right\vert
,\label{CDEF6a}\\
\nabla_{a}\left\langle \sigma,\tau\right\vert  &  =\left\langle \nabla
_{a}\sigma,\tau\right\vert +\left\langle \sigma,\nabla_{a}\tau\right\vert .
\label{CDEF6b}%
\end{align}

\item For all $\tau\in\left.  \overset{\ast}{ext}\right.  _{k}^{l}%
\mathcal{V}(U)$ and $\sigma\in\left.  \overset{\left.  {}\right.  }%
{ext}\right.  _{r}^{s}\mathcal{V}(U)$ (or, $\tau\in\left.  \overset{\left.
{}\right.  }{ext}\right.  _{k}^{l}\mathcal{V}(U)$ and $\tau\in\left.
\overset{\ast}{ext}\right.  _{r}^{s}\mathcal{V}(U)$), it holds
\begin{align}
\nabla_{a}\left\vert \tau,\sigma\right\rangle  &  =\left\vert \nabla_{a}%
\tau,\sigma\right\rangle +\left\vert \tau,\nabla_{a}\sigma\right\rangle
,\label{CDEF7a}\\
\nabla_{a}\left\vert \tau,\sigma\right\rangle  &  =\left\vert \nabla_{a}%
\tau,\sigma\right\rangle +\left\vert \tau,\nabla_{a}\sigma\right\rangle .
\label{CDEF7b}%
\end{align}

\end{itemize}

\begin{proof}
We present only the proof of the property given by Eq.(\ref{CDEF6a}). Without
loss of generality, we prove this statement only for a multiform extensor
field $\tau$ and a multivector extensor field $\sigma$ such that
$(X,\Phi)\longmapsto\tau(X,\Phi)$ and $(Y,\Psi)\longmapsto\sigma(Y,\Psi)$.%
\[%
\begin{array}
[c]{ll}%
\nabla_{a}\left\langle \tau,\sigma\left\vert \left(  X,Y,\Phi,\Psi\right)
\right.  \right.  & =\nabla_{a}\left(  \left\langle \tau,\sigma\left\vert
\left(  X,Y,\Phi,\Psi\right)  \right.  \right.  \right)  -\left\langle
\tau,\sigma\left\vert \left(  \nabla_{a}X,Y,\Phi,\Psi\right)  \right.  \right.
\\
& -\left\langle \tau,\sigma\left\vert \left(  X,\nabla_{a}Y,\Phi,\Psi\right)
\right.  \right.  -\left\langle \tau,\sigma\left\vert \left(  X,Y,\nabla
_{a}\Phi,\Psi\right)  \right.  \right. \\
& -\left\langle \tau,\sigma\left\vert \left(  X,Y,\Phi,\nabla_{a}\Psi\right)
\right.  \right.
\end{array}
\]
or recalling that $\left\langle \tau,\sigma\left\vert _{\left(  p\right)
}\right.  \right.  =\left\langle \tau_{\left(  p\right)  },\sigma_{\left(
p\right)  }\left\vert .\right.  \right.  $,
\begin{equation}%
\begin{array}
[c]{ll}%
\nabla_{a}\left\langle \tau,\sigma\left\vert \left(  X,Y,\Phi,\Psi\right)
\right.  \right.  & =\nabla_{a}\left(  \left\langle \tau\left(  X,\Phi\right)
,\sigma\left(  Y,\Psi\right)  \left\vert {}\right.  \right.  \right)
-\left\langle \tau\left(  \nabla_{a}X,\Phi\right)  ,\sigma\left(
Y,\Psi\right)  \left\vert {}\right.  \right. \\
& -\left\langle \tau\left(  X,\Phi\right)  ,\sigma\left(  \nabla_{a}%
Y,\Psi\right)  \left\vert {}\right.  \right.  -\left\langle \tau\left(
X,\nabla_{a}\Phi\right)  ,\sigma\left(  Y,\Psi\right)  \left\vert {}\right.
\right. \\
& -\left\langle \tau\left(  X,\Phi\right)  ,\sigma\left(  Y,\nabla_{a}%
\Psi\right)  \left\vert {}\right.  \right.  .
\end{array}
\label{EN1}%
\end{equation}
On the other hand, from \cite{fmcr3}, we can write
\[
\nabla_{a}\left(  \left\langle \tau\left(  X,\Phi\right)  ,\sigma\left(
Y,\Psi\right)  \left\vert {}\right.  \right.  \right)  =\left\langle
\nabla_{a}\tau\left(  X,\Phi\right)  ,\sigma\left(  Y,\Psi\right)  \left\vert
{}\right.  \right.  +\left\langle \tau\left(  X,\Phi\right)  ,\nabla_{a}%
\sigma\left(  Y,\Psi\right)  \left\vert {}\right.  \right.  ,
\]
Eq.(\ref{EN1}) can by written as%
\[%
\begin{array}
[c]{l}%
\nabla_{a}\left\langle \tau,\sigma\left\vert \left(  X,Y,\Phi,\Psi\right)
\right.  \right. \\
=\left\langle \nabla_{a}\tau\left(  X,\Phi\right)  -\tau\left(  \nabla
_{a}X,\Phi\right)  -\tau\left(  X,\nabla_{a}\Phi\right)  ,\sigma\left(
Y,\Psi\right)  \left\vert {}\right.  \right. \\
+\left\langle \tau\left(  X,\Phi\right)  ,\nabla_{a}\sigma\left(
Y,\Psi\right)  -\sigma\left(  \nabla_{a}Y,\Psi\right)  -\sigma\left(
Y,\nabla_{a}\Psi\right)  \left\vert {}\right.  \right. \\
=\left\langle \left(  \nabla_{a}\tau\right)  \left(  X,\Phi\right)
,\sigma\left(  Y,\Psi\right)  \left\vert {}\right.  \right.  +\left\langle
\tau\left(  X,\Phi\right)  ,\left(  \nabla_{a}\sigma\right)  \left(
Y,\Psi\right)  \left\vert {}\right.  \right. \\
=\left\langle \nabla_{a}\tau,\sigma\left\vert \left(  X,Y,\Phi,\Psi\right)
\right.  \right.  +\left\langle \tau,\nabla_{a}\sigma\left\vert \left(
X,Y,\Phi,\Psi\right)  \right.  \right.  ,
\end{array}
\]
which proves our result.
\end{proof}

Finally we prove that the duality adjoint operator commutes with the
$a$-\textit{DCDO}, i.e.:

\begin{itemize}
\item If $\tau$ is any one of the four smooth one-variable extensor fields on
$U$, then
\begin{equation}
(\nabla_{a}\tau)^{\bigtriangleup}=\nabla_{a}\tau^{\bigtriangleup}.
\label{CDEF8}%
\end{equation}

\end{itemize}

\begin{proof}
Without loss of generality, we prove this statement only for $\tau\in ext(%
%TCIMACRO{\dbigwedge \nolimits_{1}^{\diamond}}%
%BeginExpansion
{\displaystyle\bigwedge\nolimits_{1}^{\diamond}}
%EndExpansion
\mathcal{V}(U),%
%TCIMACRO{\dbigwedge \nolimits^{\diamond}}%
%BeginExpansion
{\displaystyle\bigwedge\nolimits^{\diamond}}
%EndExpansion
\mathcal{V}(U)).$

Let us $X\in%
%TCIMACRO{\dbigwedge \nolimits_{1}^{\diamond}}%
%BeginExpansion
{\displaystyle\bigwedge\nolimits_{1}^{\diamond}}
%EndExpansion
\mathcal{V}(U)$ and $\Phi\in%
%TCIMACRO{\dbigwedge \nolimits^{\diamond}}%
%BeginExpansion
{\displaystyle\bigwedge\nolimits^{\diamond}}
%EndExpansion
\mathcal{V}^{\ast}(U).$ We must prove that
\[
\left\langle \nabla_{a}\tau^{\bigtriangleup}(\Phi),X\right\rangle
=\left\langle \Phi,\nabla_{a}\tau(X)\right\rangle .
\]

By using Eq.(\ref{CDEF1}) and recalling the Leibniz's rule for the covariant
differentiation of the duality scalar product of multiform fields with
multivector fields, we can write
\[
\left\langle \nabla_{a}\tau^{\bigtriangleup}(\Phi),X\right\rangle
=\left\langle \nabla_{a}(\tau^{\bigtriangleup}(\Phi)),X\right\rangle
-\left\langle \tau^{\bigtriangleup}(\nabla_{a}(\Phi)),X\right\rangle ,
\]
i.e.,%
\[
\left\langle \nabla_{a}\tau^{\bigtriangleup}(\Phi),X\right\rangle
=a\left\langle \tau^{\bigtriangleup}(\Phi),X\right\rangle -\left\langle
\tau^{\bigtriangleup}(\Phi),\nabla_{a}X\right\rangle -\left\langle
\tau^{\bigtriangleup}(\nabla_{a}\Phi),X\right\rangle .
\]

Recalling now the fundamental property of the duality \textit{adjoint}, we
get
\[
\left\langle \nabla_{a}\tau^{\bigtriangleup}(\Phi),X\right\rangle
=a\left\langle \Phi,\tau(X)\right\rangle -\left\langle \Phi,\tau(\nabla
_{a}X)\right\rangle -\left\langle \nabla_{a}\Phi,\tau(X)\right\rangle .
\]
Using once again the Leibniz's rule for the covariant differentiation of the
duality scalar product we get%
\[
\left\langle \nabla_{a}\tau^{\bigtriangleup}(\Phi),X\right\rangle
=\left\langle \Phi,\nabla_{a}(\tau(X))\right\rangle -\left\langle \Phi
,\tau(\nabla_{a}X)\right\rangle ,
\]
and thus, using once again Eq.(\ref{CDEF1}), the required result follows.
\end{proof}

\section{Deformed Covariant Derivative}

Let $\left\langle U,\Gamma\right\rangle $ be a parallelism structure on $U,$
and let $\nabla_{a}$ be the associated $a$-\textit{DCDO}. By taking an
invertible smooth extensor operator field $\lambda$ on $V\supseteq U,$ we can
construct a deformed parallelism structure on $U$, denoted in \cite{fmcr3} by
$\left\langle U,\overset{\lambda}{\Gamma}\right\rangle $, with associated
$a$-\textit{DCDO} denoted by $\overset{\lambda}{\nabla}_{a}$.

As we know (see \cite{fmcr3}), such deformed covariant derivative operator
$\overset{\lambda}{\nabla}_{a}$has the following properties: for all $X\in%
%TCIMACRO{\dbigwedge }%
%BeginExpansion
{\displaystyle\bigwedge}
%EndExpansion
\mathcal{V}(U)$, $\overset{\lambda}{\nabla}_{a}X=\underline{\lambda}%
(\nabla_{a}\underline{\lambda}^{-1}(X))$, and also for all $\Phi\in%
%TCIMACRO{\dbigwedge }%
%BeginExpansion
{\displaystyle\bigwedge}
%EndExpansion
\mathcal{V}^{\star}(U)$, $\overset{\lambda}{\nabla}_{a}\Phi=\underline
{\lambda}^{-\bigtriangleup}(\nabla_{a}\underline{\lambda}^{\bigtriangleup
}(\Phi))$.

We present now two properties for $\overset{\lambda}{\nabla}_{a}$which are
generalizations from the properties just recalled above.

\begin{itemize}
\item For all $\tau\in\left.  \overset{\left.  {}\right.  }{ext}\right.
_{k}^{l}\mathcal{V}(U):$%
\begin{equation}
\overset{\lambda}{\nabla}_{a}\tau=\underline{\lambda}\nabla_{a}\underline
{\lambda}^{-1}\tau, \label{DCD3a}%
\end{equation}
where $\underline{\lambda}^{-1}\tau$ means the action of $\underline{\lambda
}^{-1}$ on the smooth multivector extensor field $\tau,$ and $\underline
{\lambda}\nabla_{a}\underline{\lambda}^{-1}\tau$ is the action of
$\underline{\lambda}$ on the smooth multivector extensor field $\nabla
_{a}\underline{\lambda}^{-1}\tau$.

\item For all $\upsilon\in\left.  \overset{\ast}{ext}\right.  _{k}%
^{l}\mathcal{V}(U):$%
\begin{equation}
\overset{\lambda}{\nabla}_{a}\upsilon=\underline{\lambda}^{-\bigtriangleup
}\nabla_{a}\underline{\lambda}^{\bigtriangleup}\upsilon, \label{DCD3b}%
\end{equation}
where $\underline{\lambda}^{\bigtriangleup}\upsilon$ means the action of
$\underline{\lambda}^{\bigtriangleup}$ on the smooth multiform extensor field
$\upsilon,$ and $\underline{\lambda}^{-\bigtriangleup}\nabla_{a}%
\underline{\lambda}^{\bigtriangleup}\upsilon$ is the action of $\underline
{\lambda}^{-\bigtriangleup}$ on the smooth multiform extensor field
$\nabla_{a}\underline{\lambda}^{\bigtriangleup}\upsilon.$
\end{itemize}

\begin{proof}
We will prove only the property for smooth multivector extensor fields as
given by Eq.(\ref{DCD3a}). Without restrictions on generality, we can work
with a multivector extensor field $(X,\Phi)\longmapsto\tau(X,\Phi)$.

Then, using Eq.(\ref{CDEF1}) and taking into account the properties just
recalled above, we can write%
\begin{align*}
&  (\overset{\lambda}{\nabla}_{a}\tau)(X,\Phi)\\
&  =\overset{\lambda}{\nabla}_{a}\tau(X,\Phi)-\tau(\overset{\lambda}{\nabla
}_{a}X,\Phi)-\tau(X,\overset{\lambda}{\nabla}_{a}\Phi)\\
&  =\underline{\lambda}(\nabla_{a}\underline{\lambda}^{-1}\circ\tau
(X,\Phi))-\tau(\underline{\lambda}(\nabla_{a}\underline{\lambda}^{-1}%
(X)),\Phi)-\tau(X,\underline{\lambda}^{-\bigtriangleup}(\nabla_{a}%
\underline{\lambda}^{\bigtriangleup}(\Phi))),
\end{align*}
i.e.,%
\begin{align*}
&  (\underline{\lambda}^{-1}\circ\overset{\lambda}{\nabla}_{a}\tau)(X,\Phi)\\
&  =\nabla_{a}\underline{\lambda}^{-1}\circ\tau(X,\Phi)-\underline{\lambda
}^{-1}\circ\tau(\underline{\lambda}(\nabla_{a}\underline{\lambda}%
^{-1}(X)),\Phi)\\
&  -\underline{\lambda}^{-1}\circ\tau(X,\underline{\lambda}^{-\bigtriangleup
}(\nabla_{a}\underline{\lambda}^{\bigtriangleup}(\Phi))).
\end{align*}

By recalling the action of an extended operator $\underline{\lambda}^{-1}$ on
a multivector extensor $\tau,$ we get%
\begin{align*}
&  (\underline{\lambda}^{-1}\circ\overset{\lambda}{\nabla}_{a}\tau)(X,\Phi)\\
&  =\nabla_{a}\underline{\lambda}^{-1}\tau(\underline{\lambda}^{-1}%
(X),\underline{\lambda}^{\bigtriangleup}(\Phi))-\underline{\lambda}^{-1}%
\tau(\nabla_{a}\underline{\lambda}^{-1}(X)),\underline{\lambda}%
^{\bigtriangleup}(\Phi))\\
&  -\underline{\lambda}^{-1}\tau(\underline{\lambda}^{-1}(X)),\nabla
_{a}\underline{\lambda}^{\bigtriangleup}(\Phi)).
\end{align*}

Using once again Eq.(\ref{CDEF1}), we have%
\[
(\overset{\lambda}{\nabla}_{a}\tau)(X,\Phi)=\underline{\lambda}\circ
(\nabla_{a}\underline{\lambda}^{-1}\tau)(\underline{\lambda}^{-1}%
(X),\underline{\lambda}^{\bigtriangleup}(\Phi)),
\]
and finally recalling once again the action of an extended operator
$\underline{\lambda}$ on a multivector extensor $\nabla_{a}\underline{\lambda
}^{-1}\tau$, the required result follows.
\end{proof}

\section{Relative Covariant Derivative}

Let $\left\langle U_{0},\Gamma\right\rangle $ be a parallelism structure on
$U_{0}$, and let $\nabla_{a}$ be as usually its associated $a$-\textit{DCDO}.
Take any relative parallelism structure $\left\langle U,B\right\rangle $
compatible with $\left\langle U_{0},\Gamma\right\rangle $ ($U_{0}\cap
U\neq\emptyset$ ). Denote by $\partial_{a}$ its associated $a$-\textit{DCDO}.
As we know (see \cite{fmcr3}),\ there exists a well-defined smooth vector
operator field on $U_{0}\cap U$, called the\emph{\ relative connection field}
$\gamma_{a}$, which satisfies the \emph{split theorem} valid for smooth
multivector fields and for smooth multiform fields, i.e.: for all $X\in%
%TCIMACRO{\tbigwedge }%
%BeginExpansion
{\textstyle\bigwedge}
%EndExpansion
\mathcal{V}(U_{0}\cap U)$,$\nabla_{a}X=\partial_{a}X+\underset{\smile}%
{\gamma_{a}}(X)$, and for all $\Phi\in%
%TCIMACRO{\tbigwedge }%
%BeginExpansion
{\textstyle\bigwedge}
%EndExpansion
\mathcal{V}^{\star}(U_{0}\cap U)$, $\nabla_{a}\Phi=\partial_{a}\Phi
-\underset{\smile}{\gamma_{a}^{\bigtriangleup}}(\Phi)$.

We now present a split theorem for smooth multivector extensor fields and for
smooth multiform extensor fields, which are the generalizations of the
properties just recalled above.

\textbf{Theorem}. \ (i) For all $\tau\in\left.  \overset{\left.  {}\right.
}{ext}\right.  _{k}^{l}\mathcal{V}(U_{0}\cap U):$%
\begin{equation}
\nabla_{a}\tau=\partial_{a}\tau+\underset{\smile}{\gamma_{a}}\tau,
\label{RCD3a}%
\end{equation}
where $\underset{\smile}{\gamma_{a}}\tau$ means the action of $\underset
{\smile}{\gamma_{a}}$ on the smooth multivector extensor field $\tau.$

( ii) For all $\upsilon\in\left.  \overset{\ast}{ext}\right.  _{k}%
^{l}\mathcal{V}(U_{0}\cap U):$%
\begin{equation}
\nabla_{a}\upsilon=\partial_{a}\upsilon-\underset{\smile}{\gamma
_{a}^{\bigtriangleup}}\upsilon, \label{RCD3b}%
\end{equation}
where $\underset{\smile}{\gamma_{a}^{\bigtriangleup}}\upsilon$ means the
action of $\underset{\smile}{\gamma_{a}^{\bigtriangleup}}$\ on the smooth
multiform extensor field $\upsilon.$

\begin{proof}
We prove only the property for smooth multivector extensor fields, i.e.,
Eq.(\ref{RCD3a}). Without loss of generality, we check this statement for a
multivector extensor field $(X,\Phi)\longmapsto\tau(X,\Phi).$

Using Eq.(\ref{CDEF1}), and by taking into account the properties just
recalled above, we have%
\begin{align*}
(\nabla_{a}\tau)(X,\Phi)  &  =\nabla_{a}\tau(X,\Phi)-\tau(\nabla_{a}%
X,\Phi)-\tau(X,\nabla_{a}\Phi)\\
&  =\partial_{a}\tau(X,\Phi)+\underset{\smile}{\gamma_{a}}(\tau(X,\Phi))\\
&  -\tau(\partial_{a}X+\underset{\smile}{\gamma_{a}}(X),\Phi)-\tau
(X,\partial_{a}\Phi-\underset{\smile}{\gamma_{a}^{\bigtriangleup}}(\Phi))\\
&  =\partial_{a}\tau(X,\Phi)-\tau(\partial_{a}X,\Phi)-\tau(X,\partial_{a}%
\Phi)\\
&  +\underset{\smile}{\gamma_{a}}(\tau(X,\Phi))-\tau(\underset{\smile}%
{\gamma_{a}}(X),\Phi)+\tau(X,\underset{\smile}{\gamma_{a}^{\bigtriangleup}%
}(\Phi)).
\end{align*}
Then, using once again Eq.(\ref{CDEF1}) and recalling the action of a
generalized operator $\underset{\smile}{\gamma_{a}}$ on a multivector extensor
$\tau$, see \cite{fmcr1}, we get
\[
(\nabla_{a}\tau)(X,\Phi)=(\partial_{a}\tau)(X,\Phi)+(\underset{\smile}%
{\gamma_{a}}\tau)(X,\Phi),
\]
and the proposition is proved.
\end{proof}

\section{Conclusions}

Using the algebraic and analytical tools developed in previous papers
\cite{fmcr1,fmcr3,fmcr4} we presented a theory of covariant derivatives,
deformed covariant derivatives and relative covariant derivatives of extensor
fields, which are essential tools in geometric theories of the gravitational
field (and also in the Lagrangian formalism of general field theory
\cite{rodoliv2006}), as will be shown in forthcoming papers.

\end{document}